\documentclass[12pt]{book}

\usepackage[dvips]{graphicx,color}
\usepackage{makeidx,hori,zon}

\makeauthorindex

\BookTitle{Inflating Horizon of Particle Astrophysics and Cosmology}
\CopyRight{\copyright 2005 by Universal Academy Press, Inc.\ and Yamada Science Foundation}

\begin{document}

\def\omunit{{(km s$^{-1}$)/kpc}}
\def\gtrsim{ \lower .75ex \hbox{$\sim$} \llap{\raise .27ex \hbox{$>$}} }
\def\lesssim{ \lower .75ex \hbox{$\sim$} \llap{\raise .27ex \hbox{$<$}} }

\def\sun{\odot}
\def\A{{\sf A}}
\def\B{{\sf B}}
\def\Ms{\mathrm{M}_\sun}
\def\Rs{\mathrm{R}_\sun}
\def\vk{v_\mathrm{kick}}
\def\ee{{\it e}}
\newcommand\arcsec{\mbox{$^{\prime\prime}$}}

\def\figurestyle#1{\sf \small \baselineskip 14pt #1}

\pagenumbering{arabic}

\chapter{
The Origin of the Binary Pulsar J0737-3039B  }

\author{%
Tsvi Piran$^{1,2,i}$ and Nir J. Shaviv$^{1,ii}$\\
{\it 1. Racah Institute for Physics, The  University, Jersualem 91904, ISRAEL\\
2. Theoretical Astrophysics, CALTECH, Pasadena, CA 91125, USA\\
(i) tsvi@huji.ac.il ~~(ii) shaviv@phys.huji.ac.il}}

%
%
\AuthorContents{T.\ Piran and N. J. Shaviv} 

\AuthorIndex{Piran}{T.}
\AuthorIndex{Shaviv}{N.J.}

\section*{Abstract}

It is generally accepted that neutron stars form in core collapse
events that are accompanied by a supernovae (types II or Ib or
Ic). Typical progenitors are, therefore, larger than $\sim 2.1
\Ms$. We suggest \cite{PS04,PS05} that the binary pulsar
J0737-3039 provides evidence for a new formation channel: collapse
of a light progenitor. This binary pulsar J0737-3039 has several
remarkable features including among others: a very tight orbit
with a Keplerian velocity of ~600km/sec, a low eccentricity, and a
location ~50pc from the Galactic plane implying that the system
has, at high likelihood, a small (compared to Keplerian) center of
mass velocity. A significant mass loss during the formation of the
second pulsar would have lead either to an eccentric orbit or to a
large center of mass velocity or to both. Therefore, we can set a
strong upper limit on the progenitor's mass. A progenitor more
massive than 1.9$\Ms$  is ruled out (at 97\% confidence).  The
kinematically favored option is of a progenitor mass around
1.45$\Ms$. Recent evidence for a rather low velocity proper motion
supports this prediction and decreases the likelihood of the
standard (high mass progenitor) scenario. Lack of variations in
the pulses' profiles (which indicate no significant geodetic
precession) provides further support for the no kick and low
progenitor mass formation scenario.

\section{Introduction}

The remarkable binary system J0737-3039 \cite{Ref1,Ref2} was
discovered during a pulsar search carried out using a multibeam
receiver at the Parkes 64-m radio telescope in New South Whales.
It is composed of two pulsars J0737-3039\A\ and J0737-3039\B\
denoted here \A\ and \B. This discovery provided a new ideal
general relativistic laboratory. The eclipses of \A\ beyond \B\
provide a superb way to explore pulsar magnetospheres. The
relatively short life time of this system to gravitational
radiation emission has lead to a revision of the binary merger
rate in the galaxy. We have suggested, immediately following the
discovery \cite{PS04}, that the orbital parameters of this system
and its location close to the Galactic plane pose strong limits on
the origin of this binary system and on the mass of the progenitor
of the younger pulsar \B. We review those arguments here and
examine how do these predictions look now almost two years after
the discovery.

We begin by summarizing the relevant parameters of this system:
The separation, $R$, (the sum of the semi-major axes) of the
pulsars today is $8.8 \cdot 10^{10}$cm. The orbit is almost
circular and the eccentricity, $\ee$, is 0.087779.  The periods
$P_{A,B}$ and their time derivatives provide upper limits for the
life times of the pulsars: $t_A \approx 210$Myr and $t_B \approx
50$Myr. Dispersion indicates that the system is 600pc from Earth.
It is located 50pc from the Galactic plane.

The present evolution of the system is determined by gravitational
radiation emission. The separation of the pulsars and the
eccentricity decrease with time and the system will merge
approximately 85Myr from now. An interesting feature of the
evolution is that the eccentricity decays with a faster rate than
the orbital separation \cite{Ref5}. Thus an eccentric orbit
becomes first circular and then it decays.  We can also integrate
backwards in time and determine the system's parameters at the
time of its birth. We find that 50Myr ago, when the pulsar \B\ was
born, the separation was ~$10^{11}$cm. The ellipticity at that
time was only slightly larger $e \sim 0.11$ than the present one.
Thus, the system was born with a low ellipticity. The values are
not that different at 200Myr. At that time, the eccentricity was
0.14 and the separation was $1.2\cdot  10^{11}$cm. As these values
are not that different from the present ones our analysis is
insensitive to the exact age of the pulsar.

The binary B1913+16 was studied for almost 3 decades since its
discovery \cite{HT75}. Wex et al. \cite{wkr00} used the ample data
to place interesting limits on this system. For example, they
found that the natal kick must have been directed almost
perpendicular to the spin axis of the neutron star progenitor. On
the other hand, the rather large post-SN eccentricity implies that
the pre-collapse orbital separation (1.8-4.6$\Rs$) or the mass of
the progenitor (4-32$\Ms$) cannot be tightly constrained. The
system J0737-3039 is different from previously detected binary
pulsar systems in that it has a very low eccentricity, and it was
small also after the collapse.  Thus, the orbital separation
before the collapse, for example, can be constrained to have been
A = $1.0 \pm 0.1 \cdot 10^{11}cm = 1.45\pm 0.15\Rs$. As the system
J0737-3039 was much tighter than B1913+16, the orbital Keplerian
velocities (and hence all other relevant velocities) are much
larger. This would play an important role in our analysis. The
location of J0737-3039 only 50pc from the Galactic plane suggests
that the system has a small center of mass velocity. This is
another important factor.

The formation of the second pulsar is described according to
current picture by a core collapse event that involves a supernova
and mass ejection from the system. Dewi and van den Heuvel
\cite{Ref6} considered a scenario in which the progenitor star
lost most of its envelope through interaction with its companion
\A. Prior to the formation of the second pulsar, tidal interaction
between the progenitor and the neutron star has led to a circular
orbit. The lost mass took place via a common envelope phase, at
which point the companion J0737-3039A was spun up and its magnetic
field was suppressed by accretion. They estimate that the
progenitor mass was more massive than $2.3 \Ms$. According to
their estimates for lower masses the common envelope phase would
evolve too fast and would result in too small separation.  The
formation of the second pulsar is described according to the
current picture by a core collapse event that involves a supernova
and mass ejection from the system. Standard evolutionary
scenarios, lead neither to neutron star formation nor to core
collapse, from progenitors that are less massive than $2.1-2.3\Ms$
\cite{Ref7,Ref8}.

We begin with a general discussion of the orbital parameters of a
binary after an instantaneous mass ejection event, such as the one
that takes place in a supernova. We then apply these general
arguments to J0737-3039. We show that that if the formation of the
second (younger) pulsar involved significant mass ejection then
the hole system would have had have a large center of mass motion,
of the order of the large Keplerian velocity. However, with a
large center of mass motion it is unlikely to find the system in
the the Galactic plane. We use these arguments to put limits on
the mass ejection and on the progenitor's mass. These arguments
were put forwards right after the discovery of the system and they
have led us to predict that the system will have a very small
(tens of km/sec) proper motion velocity \cite{PS04}. Later on
upper limits on the velocity were found and we consider the
implications of these upper limits to our considerations. We
conclude with a discussion of the implications of these results to
stellar evolutionary scenarios and to the rate of NS mergers.

\section{Mass Ejection and Orbital Motion}

We consider first the influence of mass ejection during the
formation of the second pulsar, on the orbital motion. We consider
in the following a system that due to the tidal interaction
between the older neutron star \A\ and the progenitor of \B\ was
in circular motion prior to the second supernova. During this
supernova a mass, $\Delta m$, is ejected from the star \B\ while
the star itself becomes a neutron star. The thrust of the ejected
mass, $\Delta m$, gives a velocity, $v_{cm}$, to the center of
mass (CM) of the remaining system. In addition, the mass loss
would lead to an elliptic orbit or even to the disruption of the
system. For a spherically symmetric mass loss, $v_{cm}$ will be:
\begin{equation}
v_{cm} = \left( m_{A} \Delta m \over \left( m_A + m_B\right)^{3/2}
\left( m_A + m_{Bi} \right)^{1/2}\right) v_K \label{vcm}
\end{equation}
where $v_K \equiv \sqrt{G (m_A + m_B) / R}$ is the Keplerian
velocity of the two stars relative to each other, just after the
explosion, with $R$ being the distance between the two stars at
that time. Within the context of J0737-3039, $m_A = 1.377(5)\Ms$,
$m_B = 1.250(5) \Ms$, $\Delta m$ is, of course, unknown and
$m_{Bi} \equiv m_B + \Delta m$ is the initial mass of \B\, while
$v_K \approx 600$km/s.  With $m_A \approx m_B $, $v_{cm}$ would be
of the order of $v_K/2$ unless $\Delta m \ll m_B$.

The common explanation for a birth of a system  with a low
eccentricity (with $\Delta m \approx m_B$) is for \B\ to have had
a natal kick \cite{Ref9}. While the origin of these kicks is not
clear, independent evidence for such kicks arise from the peculiar
motion of individual pulsars \cite{hp97}. Surely, with suitable
adjustment of the kick velocity the final orbit can have a low
eccentricity and even become circular. However, in order to
decrease the eccentricity the kick velocity given to \B\ must be
opposite to its actual velocity. This can be understood
intuitively. After the mass ejection  the mass of the system and
hence the gravitational attraction between \A\ and \B\ decrease
and the system finds itself with ``too much" kinetic energy. Hence
the orbit is elliptic or even disrupted. To reduce the ellipticity
we need to slow down \B\ (there is no way to slow down \A\ at this
stage) and this can be done with a kick velocity opposite to its
original velocity.

\begin{figure}[h]
  \begin{center}
    \includegraphics[height=2.5in]{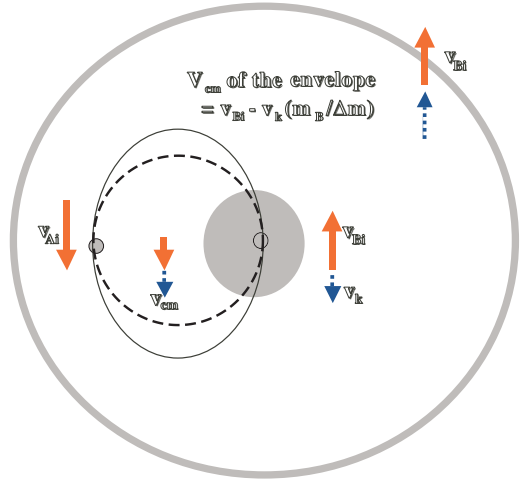}
  \end{center}
  \caption{The kinematics of mass ejection and the orbital motion with
   and without a kick velocity. The conditions
  after a kick velocity that leads to a circular orbit are marked with dotted lines.}
\label{fig_kick}
\end{figure}

Even without a kick velocity the mass ejection causes the CM of
the system to move in an opposite direction to the velocity of \B\
(see fig. \ref{fig_kick}). A kick that reduces the eccentricity
essentially increases $v_{cm}$ further and the above value can be
considered as a lower limit. In other words, the nearly circular
orbit today implies either a large CM velocity, roughly as given
by Eq. \ref{vcm}, or a small ejected mass $\Delta m \ll m_B$.

\begin{figure}[t]
  \begin{center}
    \includegraphics[height=3in]{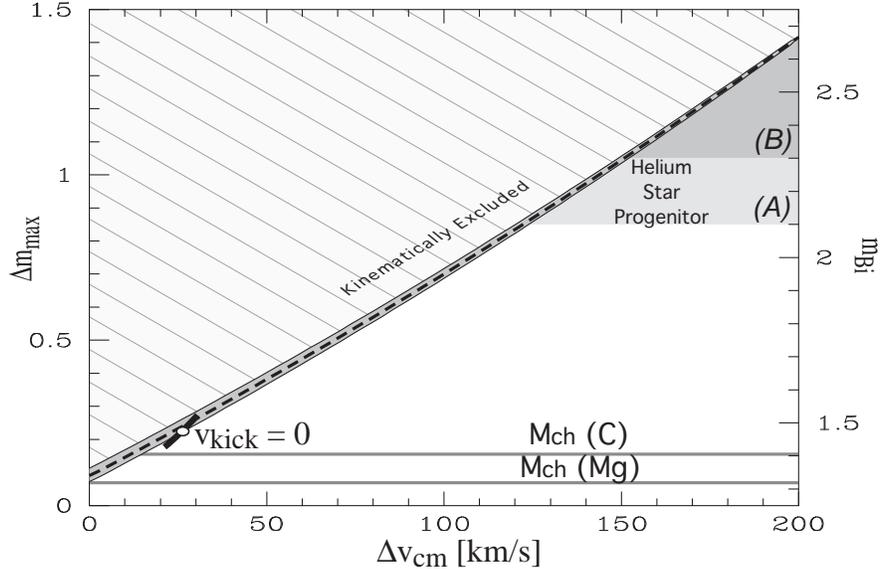}
  \end{center}
  \caption{The maximal kinematically allowed mass loss
$\Delta m_{max}$ and progenitor mass $m_{Bi}$ for a given change
in the CM velocity $\Delta v_{cm}$. The upper mass depends on the
actual eccentricity ($0.088< \ee < 0.14$), hence the finite width
of the line. The hatched area is kinematically excluded. The
region allowed by standard evolutionary scenario through the
formation of a He-star, which requires a minimum mass of $2.1\Ms$
(marked A \cite{Ref8}) to $2.3\Ms$ (marked B \cite{Ref7}) to form
a NS through a core collapse SN, is marked as shaded triangles.
The no-kick solution that results with the above range of
eccentricities is marked by the short heavy line.}
\end{figure}

Fig.\ 2 depicts change in the CM velocity (relative to the unknown
CM velocity prior to the explosion) for a system with two masses
moving on a circular orbit,  assuming the new binary system
attains the initial eccentricity of J0737-3039 ($0.088 < \ee <
0.14$). In the following we assume that this initial CM velocity
was small and we approximate $v_{cm} \approx \Delta v_{cm}$. We
will return to this point in the conclusions. Fig.\ 1 shows that
the minimal CM velocity for a $2.1\Ms$ progenitor is 120km/s. 
{The no-kick solution marked here is different from the one of
Dewi and van den Heuvel \cite{Ref6} who assume that \A\ rather
than \B\ has undergone the second collapse.}

\section{Limits on the CM Motion}

The position of the system almost in the Galactic plane sets an
additional constraint (to the one imposedby the low eccentricity)
and it has led us to predict \cite{PS04} shortly after the
discovery of the pulsar that it would have a very low CM velocity
and hence a low peculiar velocity
. Since that time there have been
several different estimates of the CM velocity of this system.
Ransom et al.\ \cite{Ref4} have estimated $v_{cm\perp}$, the CM
velocity of the binary on the plane of the sky, using the observed
scintillations of the system. They find a rather large value:
$v_{cm\perp}=141 \pm 8.5$km/s (with $96.0 \pm 3.7$ km/s along the
orbit and $103.1 \pm 7.7$km/s perpendicular to it). This value
excludes the region in Fig.\ 1 left of a vertical line of
$\sim$141km/s. These findings were questioned recently by Coles et
al. \cite{Colesetal} who suggested that the scintillation pattern
is anisotropic. When including anisotropy, they find a much lower
value: $v_{cm\perp}=66 \pm 15$km/s.   Pulsar timing give even
lower upper limit on the CM velocity of $v_{cm\perp}<30$km/s
\cite{Kramer}. The last two values are actually consistent with
each other as the estimate of 66km/s was not corrected for the
motion of the Earth \cite{Kramer}. The region in Fig.\ 1 to the
right of the vertical line of $\sim$30km/s is consistent with this
observation.

The observed distance of the system from the Galactic plane,
enables us to place a statistical upper limit on $v_{cm}$. Stars
move in a periodic motion in the vertical direction. For small
vertical oscillations, the potential of the Galaxy is harmonic:
$\Phi=2\pi G\rho_0 z^2$, where $\rho_0 \approx 0.25\Ms/$pc$^3$ is
the mass density in the disk \cite{Ref11}. This gives a vertical
orbital period, $P_z \approx 50$Myr. The typical velocity for an
object at $z_{obs}$ is $v_z \approx 2\pi z_{obs}/P_z$. $z_{obs}
\approx 50$pc implies then that the expectation value of the
vertical velocity is of the order of 6km/s.

To quantify the probability for having a particular CM velocity
given the observed $z_{obs}$ we perform  Monte Carlo simulations
that follow the formation of the system. We assume  that star \B\
had a given mass $m_{Bi}$, and that a randomly oriented kick ${\bf
v}_{kick}$ was given to it. We also assume that the progenitor
distribution has an initial Gaussian distribution in the amplitude
of the vertical oscillation, with a width $\sigma_z = 50$pc (other
$\sigma_z \lesssim 100$pc gave very similar results). At the
moment of formation, we assume it had a random phase within its
vertical motion. We calculate the CM kick velocity ${\bf v}_{cm}$,
and assign it a random direction, then integrate the vertical
motion of the pulsar for $50$Myr using a realistic Galactic
potential \cite{Ref12}.

\begin{figure}[h]
  \begin{center}
    \includegraphics[width=3.4in]{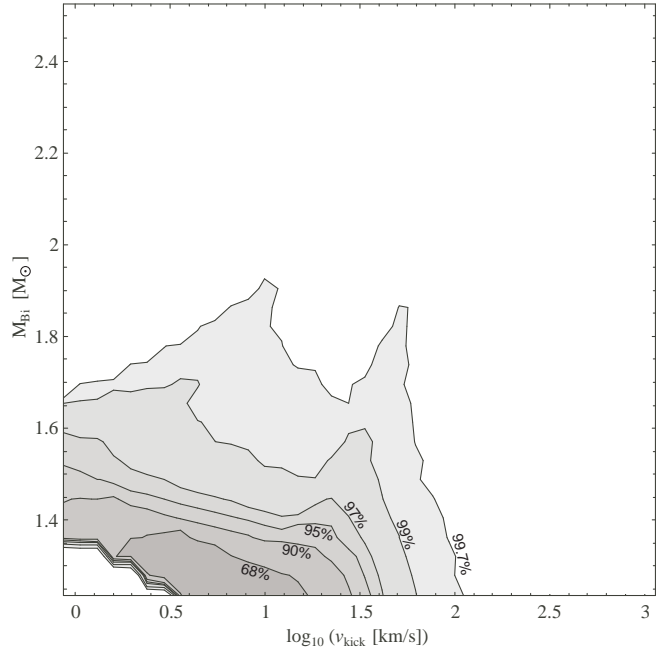}
  \end{center}
  \caption{The probability that a binary system will end up within $50$pc
from the Galactic plane, with $0.088 <\ee < 0.14$ and with a
transverse velocity less than $30$ km/s, given that $50$Myrs
before, the progenitor system had a circular orbit, that star \B\
had a progenitor mass $m_{B,i}$, and it obtained a random kick
velocity of size $\vk$. The probabilities are normalized to the
most likely $M_{Bi}, \vk$. The He-star solution requires $m_{Bi}$
larger than about $2.1\Ms$. With a fine tuned $\vk$, this type of
a solution is ruled out at the 99.7\% c.l. Solutions with $m_{Bi}
\approx 1.5 \pm 0.2\Ms$ and $\vk \lesssim 30$km/s are
kinematically more favorable.}
\end{figure}

Fig.\ 3 depicts the probability that a system with a given
$m_{Bi}$ and $v_{kick}$ could find itself with $0.087<  \ee< 0.14$
after $50$Myr, within $50$pc of the Galactic plane and with a
transverse velocity less than  $30$km/s as given by pulsar timing
\cite{Kramer}.  Other probability distributions with different
assumptions on the current status of the system can be seen in
\cite{PS05}. With the constraint that the transverse velocity is
less than $30$ km/s,  a fine tuned $m_{Bi}=2.1\Ms$ model can be
ruled out at even 99.7\% c.l. Even without this constraint on the
velocity, without any information on the CM velocity, we find
\cite{PS05} that even if $v_\mathrm{kick}$ is fine tuned to be
near either 130 or 315km/s, a system with a mass of
$m_{Bi}=2.1\Ms$ could result with the observed configuration in
only about 3\% of the random realizations (as compared with the
most favorable conditions having lower $m_{Bi}$ and
$v_\mathrm{kick}$).  Other $\vk$'s, or higher mass systems are
kinematically even less likely. On the other hand, low ejected
mass solutions are favored.  This was the basic argument of
\cite{PS04}.

An inspection of Fig.\ 3 reveals that there is a kinematically
favorable solution with a small mass loss and natal kicks ranging
from 0 to $\sim 50$km/s. Since a small mass loss necessarily
implies a small natal kick, the solutions with $v_{kick} \gtrsim
50$km/s are physically unlikely. If we therefore limit ourselves
to $v_{kick}\lesssim 30$km/s, then without fine tuning, a large
fraction of the progenitor systems will result with the observed
configuration with a progenitor mass of $1.45\Ms \lesssim m_{Bi}
\lesssim 1.65\Ms$. Thus, a mass loss of about $0.3 \pm 0.1\Ms$ is
most probable. The results are qualitatively the same even if we
do not enforce that the resulting system has a CM velocity of
66km/s, or if we replace the condition on the eccentricity to $\ee
< 0.14$ rather than $ 0.088 <\ee <0.14$.

Willems and Kalogera \cite{Ref10} estimated the allowed $\Delta m$
and $v_{kick}$ assuming, following standard scenarios,
$2.1<m_{Bi}< 4.7\Ms$. They  find $50 < v_{kick}< 1560$km/s and a
most likely value of 150km/s. However, in their analysis Willems
and Kalogera implicitly assume that the current radial CM velocity
of the system (which has not been measured) is large, impliying
that the system is moving with a very large velocity almost
directly towards us.  In fact their assumption on the magnitude of
this velocity dictates their resulting probability distribution of
the kick velocities. Our calculations include the geometric
probability that a large CM motion will be directed towards us.
When this factor is included we find, as explained above that the
``standard" high mass progenitor and large kick scenario is ruled
out.

\section{Conclusions and Implications}

We are left with two physically distinguished scenarios for the
formation of pulsar \B. In the first, the progenitor is a
kinematically ``unlikely" but theoretically plausible $2.1-
2.3\Ms$ He-star progenitor---around the minimal masses estimated
from stellar evolution scenario \cite{Ref6,Ref7,Ref8}. In the
second scenario, the progenitor is a kinematically favorable $\sim
1.5\Ms$ young stellar core. The more probable low mass solution
requires a new type of stellar collapse. Intermediate solutions
are neither statistically favored nor do they fit any plausible
theoretical scenario.

Our conclusions were based on the low eccentricity of the observed
system and on its location near the Galactic plane. When we first
suggested this model in January 2004 the CM velocity of the system
was not known and we have predicted a low CM velocity \cite{PS04}.
This was confirmed later by pulsar timing observations
\cite{Kramer} that have shown that the system has a very low
peculiar motion. An independent supportive evidence for out low
(or no) kick model is given by the fact that pulsar \A\ did not
show variations in its pulse shape  in fifteen months of
observations \cite{Kramer}. Such variations were expected
\cite{JR04} due to geodetic precession. While this observations
may results from a particular precession phase, it is more likely
that it arises due to alignment of the spin of pulsar \A\ and the
orbital angular momentum. This  suggests, in turn, that the
orbital angular momentum hasn't changed in the second collapse and
indicates a low kick velocity during the formation of
\B\footnote{A large kick velocity would presumably have a
component out of the original orbital plane and would have changed
the direction of the orbital angular momentum. }.

Before turning to the implications of these two solutions we
consider, first, three assumptions made in our analysis. (i) We
have assumed that prior to the formation of the second pulsar the
system was in a circular motion. This  follows from all
evolutionary scenarios that lead to a neutron star and a small
mass progenitor (even a $2.3 \Ms$ is a very small mass
progenitor). (ii) We have assumed that the second mass loss was
instantaneous, namely, shorter  than a fraction of an orbital
period. Given that the orbital motion is of several hundred km/s
while typical mass ejection velocities in SNe are higher than
10,000km/s, this assumption is reasonable (for all conventional
neutron star formation scenarios). (iii) We have assumed that the
CM velocity, prior to the formation of the second pulsar was
small. One would expect that the system would have acquired a CM
velocity during the formation of the first pulsar. About half of
the system's mass was lost during this event and this should have
resulted in some CM velocity. However, this velocity would have
been of the order of half the Keplerian velocity of the system at
that time (see Eq. \ref{vcm}). Given the fact that the orbital
separation was much larger we could reasonably expect, but not
prove, that this velocity was of order of several tens of km/s.
Moreover, if the system would have acquired a large CM velocity in
the first SN, there would have been an even smaller probability to
find it in the Galactic plane today.

We turn now to the most likely scenario, the very low mass
scenario. We imagine the same evolutionary scenario in which some
time before 50 Myr, system \A\ and progenitor \B\ were in a common
envelope phase and \B\ lost most of its mass keeping practically
just its core of $\sim 1.45\Ms$. This progenitor leads to a small
CM motion and does not require any kick velocity. This solution is
kinematical preferred. However, it requires a new mechanism for
the formation of the pulsar as He-stars of $1.45\Ms$ do not
collapse to form neutron stars \cite{Ref7,Ref8}.  The observation
that the progenitor mass is very close to the Chandrasekhar mass
leads us to conjecture that the process involves the collapse of a
supercritical white dwarf. For example, the progenitor may have
been a degenerate bare core just above the critical Chandrasekhar
mass, supported by the extra thermal pressure against collapse. As
it cooled, the additional support was lost and the core collapsed
to form a neutron star. A second possibility is that it was formed
just below the Chandrasekhar mass, and as it cooled,
neutronization at the core increased the baryon to electron ratio,
and with it reduced the Chandrasekhar mass until the progenitor
became unstable and collapsed. Note that the object must have been
composed of O-Ne-Mg as a collapsing CO core would have
carbon-detonated and it would have exploded completely forming a
type I SNe and leaving no remnant.  However, it seems that the
parameters space for these two  scenarios is quite small. A second
possible formation scenario is that the progenitor of \B\ had a
core of about $1.4 \Ms$ and it was gradually stripped of its
envelope until it eventually collapsed as a bare core with almost
no mass ejection. In either case it is clear that whatever the
formation scenario was it is drastically different from the
``standard" scenarios for NS formation.

This new solution passes an immediate non trivial test. With a
small mass loss, only small kick velocities are possible and in
fact we can estimate in this case (see Fig.\ 3) the mass loss
needed to obtain the initial eccentricity $e \approx 0.11$. We
find (while conservatively assuming that the collapse took place
$50^{+100}_{-50}$Myr ago) $\Delta m = e(1+q) m_B = 0.28 \pm
0.07\Ms$  and a progenitor mass of $1.53\pm 0.07 \Ms$, which is,
indeed, just above the Chandrasekhar limit. Intriguingly, some
mass loss, in the form of $\nu$ losses of a few times
$10^{53}$ergs, must take place. The estimated $\Delta m$
corresponds to $E_\nu \approx \Delta mc^2 \approx 4.2 \cdot
10^{53}$ergs, which is in the right range. Of course, if some mass
is ejected as well, $E_\nu$ will be smaller and a small kick could
arise, but the total mass-energy lost will be the same. The
consistency of this mass and energy loss with the previous
physical picture increases our belief in this new and unusual
scenario.

It is intriguing that (after the publication of our first
suggestion \cite{PS04}) other arguments has led  Van den Heuvel
\cite{vdH04} to conclude that about half of the pulsars that are
in binary systems had low mass progenitors.

\def\mnras{Mon.\ Not.\ Roy.\ Astr.\ Soc.}
\def\apj{Ap.\ J.}
\def\apjl{Ap.\ J.}

\end{document}